\documentstyle[emulateapj,10pt,apjfonts,psfig]{article}
\def\snr{G292.0+1.8}
\def\psr{J1124--5916}
\def\cxo{{\em Chandra}}
\def\cala{PKS~B1105--680}
\def\calb{PKS~B1148--671}

\newcommand\HI{H\,{\sc i}}

\def\kms{km~s$^{-1}$}
\def\etal{{\rm et~al.\ }}

\lefthead{GAENSLER \& WALLACE}
\righthead{RADIO STUDY OF SNR~\snr}

\begin{document}
\title{A Multi-Frequency Radio Study of Supernova Remnant \snr\
\\ and its Pulsar Wind Nebula}
\submitted{Accepted to {\em The Astrophysical Journal} on 09-May-2003}
\author{B. M. Gaensler\altaffilmark{1} and B. J. Wallace\altaffilmark{2}}
\altaffiltext{1}{Harvard-Smithsonian
Center for Astrophysics, 60 Garden Street MS-6, Cambridge, MA 02138;
bgaensler@cfa.harvard.edu}
\altaffiltext{2}{Defence R\&D Canada --- Ottawa,
3701 Carling Avenue,
Ottawa, ON, K1A 0Z4, Canada; Brad.Wallace@drdc-rddc.gc.ca}
\hspace{-5cm}

\begin{abstract}

We present a detailed radio study of the young supernova remnant (SNR)
\snr\ and its associated pulsar PSR~\psr, 
using the Australia Telescope Compact Array at observing wavelengths
of 20, 13 and 6~cm. We find that the radio morphology of the source
consists of three main components: a polarized flat-spectrum central core
coincident with the pulsar \psr, a surrounding circular steep-spectrum
plateau with sharp outer edges and, superimposed on the plateau, a
series of radial filaments with spectra significantly flatter than
their surroundings. \HI\ absorption argues for a lower limit on the
distance to the system of 6~kpc.

The core clearly corresponds to radio emission from a pulsar wind
nebula powered by PSR~\psr, while we conclude that the plateau
represents the surrounding SNR shell. The plateau's sharp outer rim
delineates the SNR's forward shock, while the thickness of the plateau
region demonstrates that the forward and reverse shocks are
well-separated. Assuming a distance of 6~kpc and an age for the source
of 2500~yr, we infer an expansion velocity for the SNR of
$\sim1200$~\kms, an ambient density $\sim0.9$~cm$^{-3}$, an ejected
mass $\sim5.9$~M$_\odot$ and a supernova explosion energy
$\sim1.1\times10^{51}$~erg.  We interpret the flat-spectrum radial
filaments superimposed on the steeper-spectrum plateau
as Rayleigh-Taylor unstable regions between the forward and
reverse shocks of the SNR. The flat radio spectrum seen for these
features results from efficient second-order Fermi acceleration in
strongly amplified magnetic fields. Overall, SNR~\snr\
shows an unusual set of properties not seen in any other SNR.
This source may reflect a unique stage in evolution,
only seen for systems at an age of $\sim2500$~yr,
and only for which there is both a bright SNR shell
and an energetic associated pulsar.

\end{abstract}


\section{Introduction}
\label{sec_intro}

When a massive star ends its life in a supernova explosion, the
expectation has been that this should leave behind an expanding supernova
remnant (SNR) with oxygen-rich ejecta, a central neutron star observable
as a radio pulsar, and a synchrotron-emitting pulsar wind nebula (PWN)
powered by the rotating neutron star.  However, as yet only one system
in the Galaxy has been identified which has all these properties: the
SNR~\snr\ (also known as MSH~11--5{\em 4}) and its associated pulsar,
PSR~\psr.  These sources thus present an ideal opportunity to study the
aftermath of a core-collapse supernova, and the resulting interaction
of its products with the interstellar medium (ISM) and with each other.

\snr\ was first detected through its radio emission (\cite{msh61}).  On
the basis of its non-thermal radio spectrum and lack of recombination
lines (\cite{mil69}; \cite{wmgm70}), \snr\ was subsequently classified
as a SNR, while \HI\ absorption showed it to be at a distance of at least
3.7~kpc (\cite{cmr+75}). The first sub-arcminute resolution image of
the SNR revealed a filled-center radio morphology (\cite{lgcm77}),
which led to its classification as a ``Crab-like'' SNR, i.e.\ its
emission was interpreted as being dominated by that from a PWN, rather
than from the surrounding SNR ejecta and their interaction with the
ISM. However, more sensitive radio observations demonstrated the SNR
to be composed of both a bright central ridge and a surrounding plateau
(\cite{bgcr86}).  This led Braun \etal\ (1986\nocite{bgcr86}) to
reinterpret the SNR as a thick shell interacting with dense material,
with no pulsar-powered component.

The classification of \snr\ as a ``shell-type'' SNR was supported
by observations in other wavebands. In the optical, high-velocity
oxygen rich filaments were identified, characteristic of emission from
a young SNR with a massive progenitor (\cite{gsz+79}; \cite{mc79};
\cite{van79}; \cite{bgdb83}), and arguing for an age $\sim1700$~yr.
In X-rays, the SNR showed an ellipsoidal disk of emission (\cite{tcb82}),
with an emission-line spectrum resulting from the shock-heated ejecta
of a massive progenitor (\cite{ctb80}; \cite{hs94b}).  The SNR also
showed extensive infra-red emission resulting from shock-heated dust
(\cite{bgcr86}). Until a few years ago, it thus seemed clear that
SNR~\snr\ was a shell SNR, with no known central pulsar or PWN.

New telescopes with imaging capabilities in hard X-rays have revised
this classification. {\em ASCA}\ observations of this source lacked high
spatial resolution, but clearly showed there to be a hard, compact X-ray
source embedded in extended softer emission from the rest of the SNR
(\cite{tts98}) --- this region of harder emission was interpreted as X-ray
synchrotron emission from a central PWN. High-resolution observations
of SNR~\snr\ with the {\em Chandra X-ray Observatory}\ dramatically
confirmed this claim, demonstrating
the presence of a filled-center X-ray PWN, $1'-2'$ across,
slightly offset from the center of the surrounding SNR (\cite{hsb+01}).
These observations also revealed a point-like source within the PWN,
which Hughes \etal\ (2001\nocite{hsb+01}) argued was the central powering
pulsar. A deep search for pulsations at this position, using the Parkes
radio telescope, has indeed resulted in the detection of a radio pulsar,
PSR~\psr, at this position (\cite{cmg+02}); the pulsar has a spin-period
of 135~ms, a spin-down luminosity of $1.2\times10^{37}$~erg~s$^{-1}$
and a characteristic age of 2900~yr, properties all consistent with it
being the young neutron star associated with the coincident SNR.
Pulsations have since been detected from this source in X-rays
also (\cite{hs03}).

It is thus now clear that SNR~\snr\ is a ``composite'' SNR, clearly
showing a SNR blast-wave, a central pulsar and a PWN.  We here present
an extensive multi-frequency study of the radio emission from \snr\ using
the Australia Telescope Compact Array (ATCA), the observations for which
are of far higher angular resolution and sensitivity, and the spectral
coverage much broader, than any existing radio studies of this source.
The aims of this experiment are to properly characterize the morphological
and spectral properties of the radio emission from the SNR and its PWN, to
constrain the distance to the source, and to compare the radio and X-ray
properties of the SNR. In \S\ref{sec_obs} we summarize our observations
and analysis of the radio data, in \S\ref{sec_results} we describe the
resulting images, \HI\ absorption and spectral index distribution for this
source, and in \S\ref{sec_discuss} we interpret these data in the context
of the distance to the source, and the properties of the SNR and PWN.

\section{Observations and Analysis}
\label{sec_obs}

Observations of SNR~\snr\ were carried out with the ATCA
(\cite{fbw92}), a six-element synthesis telescope located near
Narrabri, NSW, Australia.  These observing runs utilized a variety of
array configurations and observing frequencies, summarized in
Table~\ref{tab_obs}. Each observation consisted of a dual-frequency
12-hour synthesis.
Continuum observations were carried out at wavelengths of
20~cm (center frequencies of 1375, 1344 and 1472~MHz), 13~cm (2240 and
2496~MHz) and 6~cm (4800, 5056, 5312 and 5440~MHz), as listed in
Table~\ref{tab_obs}; multiple center frequencies were used within each
waveband to improve the $u-v$ coverage of the observations. All
continuum observations used 32 channels recorded across a 128-MHz
continuum band, and included data in all four Stokes parameters.
Spectral line observations were carried out centered on the 1420-MHz
\HI\ line, and consisted of 1024 channels across a 4-MHz band; only total
intensity was recorded in these data.  Observations of \snr\ at 20 and
13~cm, and in the \HI\ line, were carried out in a single pointing
offset from the SNR by a few arcmin, so as to prevent artifacts at the
phase center from corrupting the image. Observations at 6~cm consisted
of a 3-point mosaic centered on the SNR. The shortest array spacing
in each waveband was 31~meters, corresponding to a maximum
angular scale to which the array was sensitive of $\sim24'$ and $\sim14'$
at 20 and 13~cm respectively. At 6~cm, the mosaicing process
allowed us to recover somewhat shorter spacings (\cite{er78}), so that we can
image scales as large as $\sim10'$. In all cases, the array
was sensitive to all scales on which emission 
is expected to be produced by the SNR, down to the resolution limit.

Flux calibration for all observations was carried out using
data taken on PKS~B1934--638.\footnote{see {\tt
http://www.narrabri.atnf.csiro.au/observing/users\_guide/html/node215.html}}
Antenna gains and polarization calibration were measured using
regular observations of either \cala\ or \calb, as indicated in
Table~\ref{tab_obs}.
Data were edited and calibrated using the MIRIAD package (\cite{sk99}).
During this reduction, it became apparent that the 20-cm observations
of \cala\ contained significant extended structure. To determine the
time-variation in the complex antenna gains we used self-calibration to
produce a high-fidelity image of this calibration field, and then applied the
resulting gain solutions to \snr.

Total intensity images at 20, 13 and 6 cm were formed using
multi-frequency synthesis to maximize $u-v$ coverage, and with uniform
visibility weighting so as to minimize sidelobes. The 20-cm image was
formed from all available data, while the images at 13 and 6~cm
excluded all baselines longer than 3~km so as to boost signal-to-noise
and to match the resolution of the 20-cm data. 
Each image was deconvolved using a maximum entropy algorithm; the
6-cm mosaic fields were all deconvolved simultaneously using the MOSMEM
algorithm (\cite{ssb96}). Each image was then smoothed with a gaussian
restoring beam (the dimensions of which are listed in
Table~\ref{tab_src}), and then corrected for the mean primary
beam response of the antennas.

A compact source embedded in \snr, at position (J2000)
RA $11^{\rm h}24^{\rm m}25\fs19$, Dec $-59^{\circ}13'46\farcs4$,
appears to have varied in its flux density between epochs and
consequently produces sidelobes which corrupt the 
images. At each epoch, we have consequently subtracted the 
Fourier transform of this source out of the $u-v$ data before
making the image from the combined data-sets. Once these
data had been successfully deconvolved and restored, we added
back to the image an unresolved source at this position
with the appropriate mean flux density. 

Images in Stokes~$Q$, $U$ and $V$ were similarly formed in each
waveband.  At 20 and 13~cm these images were deconvolved using the
CLEAN algorithm, while at 6~cm we used the PMOSMEM algorithm
(\cite{sbd99}).  $Q$ and $U$ images in each waveband were then combined
to form images of polarized intensity and polarized position angle; a
correction for the Ricean bias was applied to these data. Examination
of the polarized position angles on a channel-by-channel basis within
the 13~cm band (see \cite{gmg98}) demonstrated that the rotation
measure (RM) towards the SNR is low, typically $|{\rm RM}| \la
30$~rad~m$^{-2}$. There is thus only $\sim20^\circ$ of Faraday rotation
between the 13 and 6~cm data.  We thus directly compared these two
data-sets to correct for Faraday rotation and derive the intrinsic
polarized position angles, without concern as to possible ambiguities
between wavebands.

To properly compare interferometric data-sets at different wavelengths
and thus determine spectral indices, one must first ensure that the
corresponding images are matched in $u-v$ coverage. We thus re-sampled
the 20~cm observations so that they matched the 13~cm and 6~cm data in
their $u-v$ sampling, primary beam shape and mosaicing pattern, and
similarly re-sampled the 13~cm data to match the 6~cm data (see
\cite{gbm+99}).
The re-processed data-sets were then imaged and deconvolved in
identical fashion as to the data to which they had been matched.  This
produced three pairs of images (20/13, 20/6 and 13/6) which were
identical in their spatial sampling and fidelity, and differed only in
their brightness distributions. These images could then be directly
compared.

For the \HI\ data, we subtracted the continuum contribution to
the visibilities in the $u-v$ plane using the UVLIN algorithm
(\cite{sau94}). Cubes
of emission were then formed for each spectral line, in
which the 1024 visibility channels were rebinned to form 207
channels each of velocity width 1.7~\kms, ranging between --100 and
+250~\kms. In the \HI\ line, we enhanced our surface-brightness
sensitivity by only using baselines shorter than 1200~m to make our
images. The resulting cubes were deconvolved using 1000 iterations of
the CLEAN algorithm, and then restored with a gaussian beam. In the
\HI\ line, we extracted absorption spectra by weighting the cube by the
20-cm continuum emission from the same region, averaging the spectrum
over a particular region of interest, and then normalizing
appropriately to give units of fractional absorption. We determined the
absorption threshold above which a feature is regarded as significant
by measuring the rms fluctuations in line-free channels, and adopting a
value six times higher than this as our sensitivity limit.  We adopt a
factor of six to reflect the increased system temperature in the
\HI\ line (e.g.\ \cite{dic97}).

\section{Results}
\label{sec_results}

\subsection{Imaging and Radio/X-ray Comparison}

\begin{figure}[hbt]
\centerline{\psfig{file=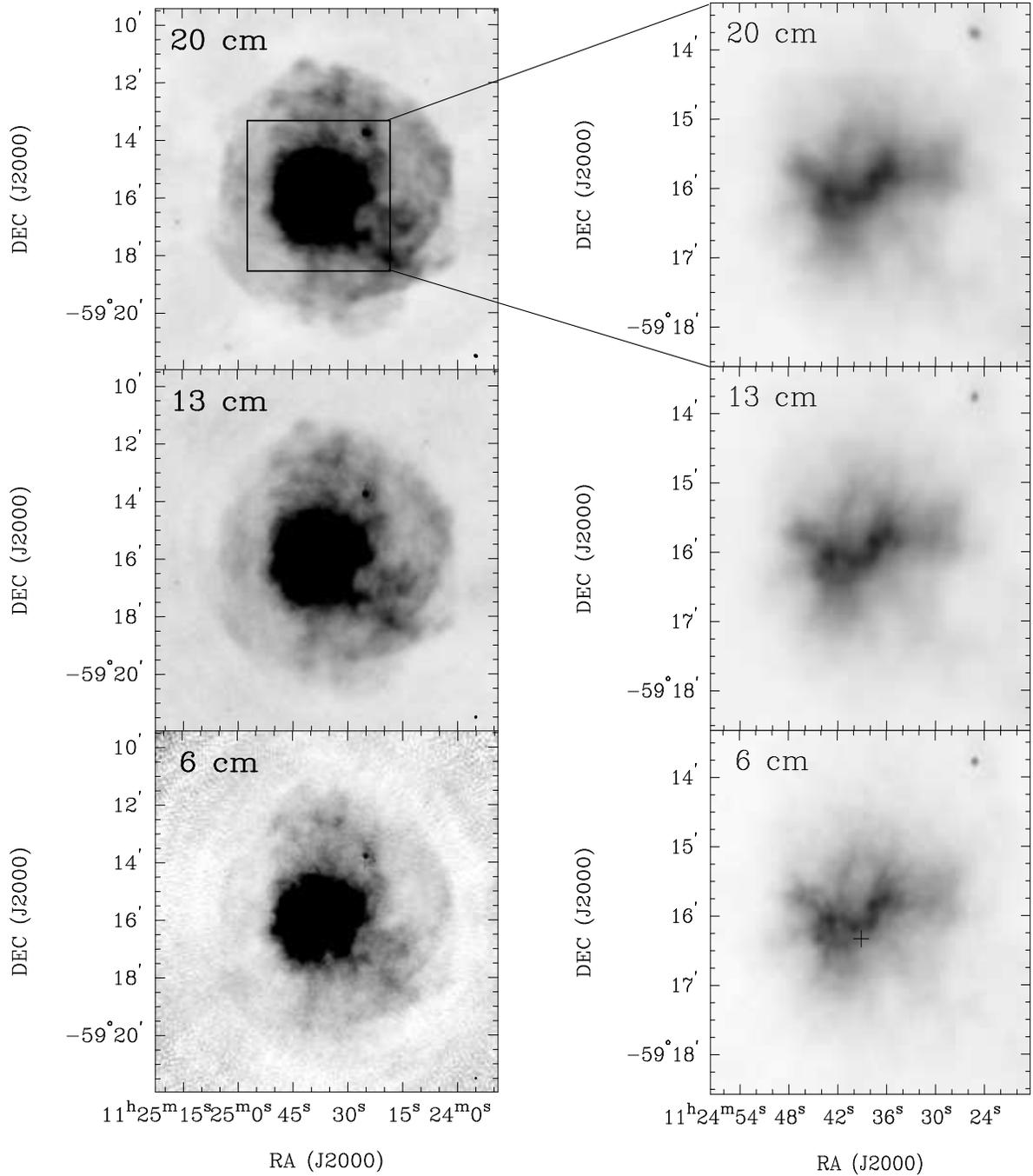,height=18cm}}
\caption{Radio continuum images of SNR~\snr\ at 20, 13 and 6~cm;
the left set of images shows the entire source, while the right
set shows the inner core. The ``+'' symbol in the 6-cm image
of the core marks the position of PSR~\psr.
Each image has been corrected for primary beam attenuation;
the dimensions of the synthesized beam are shown in
lower right corner of the wider-field images.
The greyscale ranges in the left set of
images are --0.05 to +0.5~Jy~arcmin$^{-2}$, and
in the right set of images are --0.09 to +3.5~Jy~arcmin$^{-2}$.}
\label{fig_all}
\end{figure}

\begin{figure}[hbt]
\centerline{\psfig{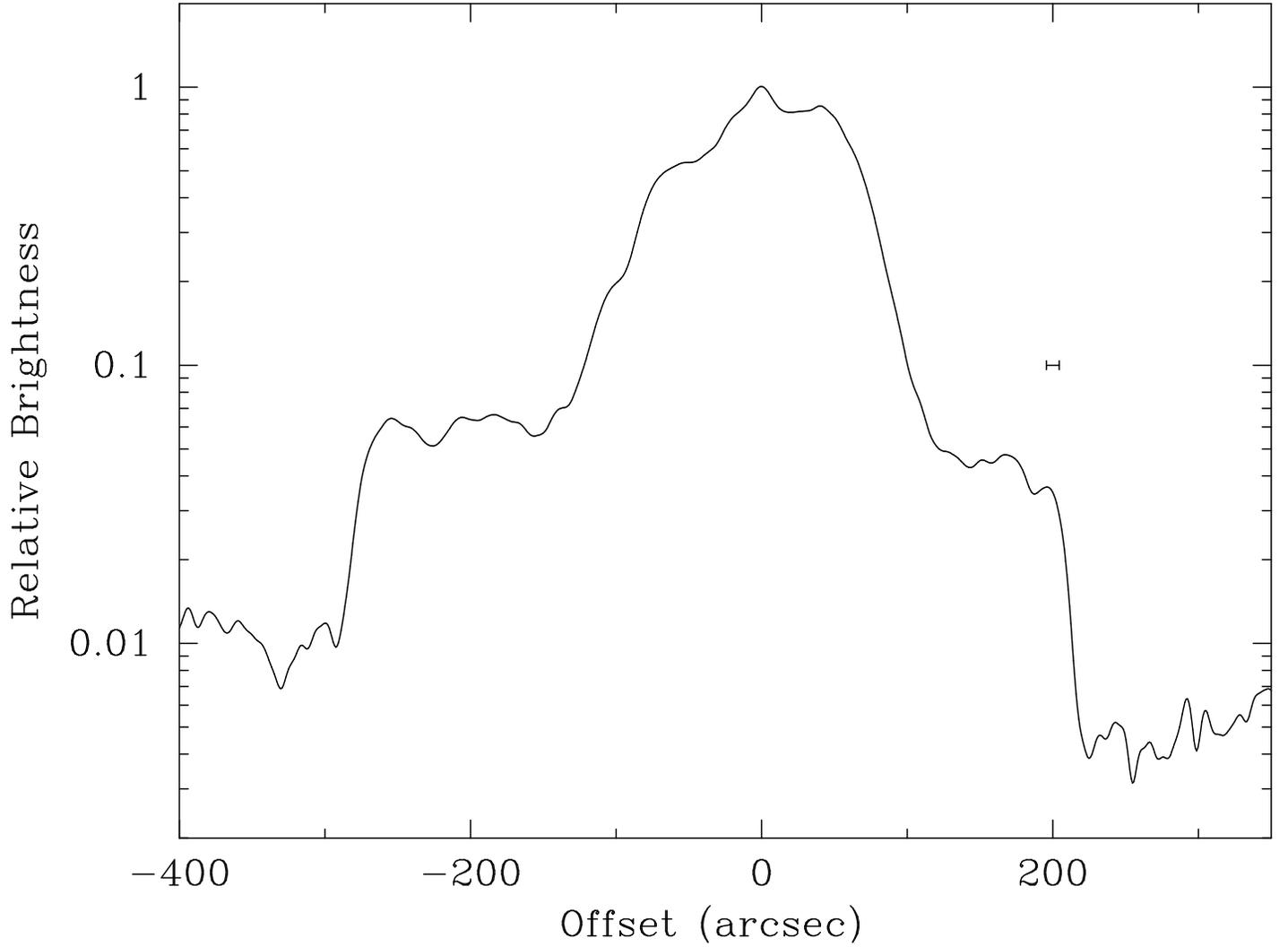}}
\caption{Profile of the 20-cm surface brightness of SNR~\snr. The
data plotted represent an east-west slice through the SNR
at a declination of (J2000) $-59^\circ 15' 55\farcs8$;
the x-axis corresponds to the relative offset from the brightest
pixel along this slice (positive offsets are in a westerly direction).
A slight offset has been added to the surface
brightness value of each datum to ensure that all values
are positive before taking their logarithm. The
error bar shows the FWHM of the synthesized beam.}
\label{fig_profile}
\end{figure}

Radio images of \snr\ at 20, 13 and 6~cm (without any filtering applied
to match the range of spatial scales measured in each case) are shown
in Figure~\ref{fig_all}, while a slice through the SNR is shown in
Figure~\ref{fig_profile}.  These data clearly demonstrate that the
morphology of \snr\ is comprised of at least two components, 
confirming similar claims made from lower resolution data
(\cite{lgcm77}; \cite{bgcr86}).  Concentric with the SNR center,
approximately circular, and with an angular diameter of $4'$, is a
bright centrally-filled region, which we follow previous authors in
referring to as the ``core''.  PSR~\psr\ is within this region, offset
by $\sim30''$ to the south-east of the core's center. The brightest
part of the core has a ridge-like morphology, primarily elongated
east-west. As shown in the lower-right panel of Figure~\ref{fig_all},
this ridge runs just to the north of the pulsar.

Surrounding the core is a fainter region of diameter $8'$.
This region, which Lockhart \etal\ (1977\nocite{lgcm77}) designated 
the ``plateau'', is somewhat brighter
on its western half compared to the eastern side of the source. 
Figure~\ref{fig_profile} demonstrates that the
plateau has sharp outer edges, over which the surface brightness drops
to that of the background in less than a single beamwidth. The
circumference of the plateau in some places shows several straight-edged
segments, 
rather than a single curved perimeter. This is most striking along the
eastern and north-eastern edges of the source.

To determine the flux density for the entire SNR, we measured the flux
within a circle centered on RA (J2000) $11^{\rm h}24^{\rm m}34\fs2$,
Dec.\ (J2000) $-59^{\circ}15'54''$ with radius $288''$. We determined
the background for this region by measuring the flux in an annulus of
the same center, and with inner/outer radii $288''$/$370''$. We
similarly measured the flux density of the core region, using a circle
and annulus centered on RA (J2000) $11^{\rm h}24^{\rm m}37\fs6$, Dec.\
(J2000)
$-59^{\circ}15'55''$, with circle radius $126''$, and annular
inner/outer radii $126''$/$164''$. The flux for the plateau is then
that of the whole source less that of the core.  The resulting
measurements at each wavelength are given in Table~\ref{tab_src}.

A number of smaller-scale radio features can be seen super-imposed on
the plateau region. Most prominent are two spurs of emission south-west
of the center; fainter radial filaments can be seen to the north and
south of center.  The presence of these structures at all three
wavelengths demonstrates that they are real features of the source
rather than artifacts of the imaging and deconvolution process.

\begin{figure}[hbt]
\centerline{\psfig{file=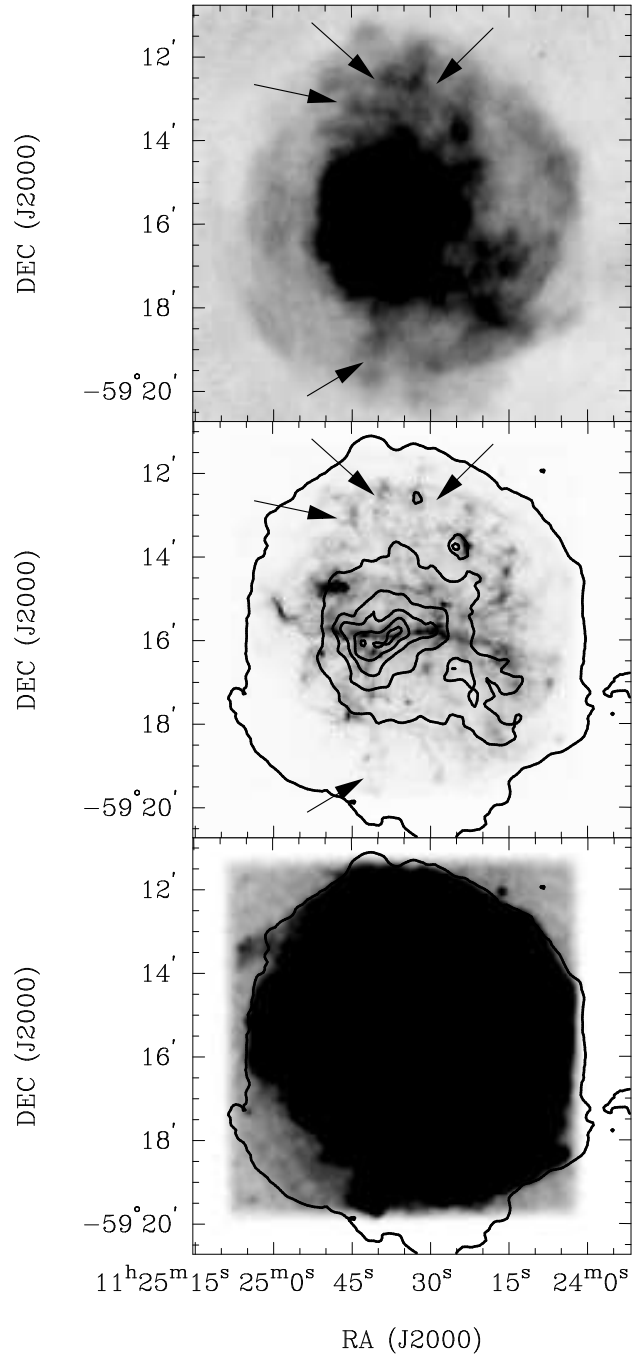,height=18cm}}
\caption{Comparison of X-ray and radio morphologies of SNR~\snr.  The
image in the upper panel corresponds to 13-cm ATCA data, while the lower two
panels show the \cxo\ ACIS-S data of Park
\etal\ (2002\protect\nocite{prh+02}) in the energy range
0.3--10~keV. The center
panel shows the morphology of the brightest X-ray emission,
overlaid with 20-cm radio contours at the levels of  0.5, 5, 15, 25, 35 and
45~mJy~beam$^{-1}$.
The bottom panel has been smoothed to a resolution of $10''$
and then heavily saturated to show the faintest structure;
the single contour plotted represents 20-cm ATCA data
at a level of 0.5~mJy~beam$^{-1}$. The arrows in the upper
two panels indicate matching radio/X-ray filaments to
the north and south of the core region.}
\label{fig_chandra}
\end{figure}

Figure~\ref{fig_chandra} compares our ATCA observations with the {\em
Chandra}\ data of Park \etal\ (2002\nocite{prh+02}).  It can be seen
from the lower two panels of this comparison that while all the bright
X-ray emission is confined to a circular region well within the radio
perimeter of the plateau, there is, as noted by Park
\etal\ (2002\nocite{prh+02}), diffuse X-ray emission at larger radii,
at a surface brightness $\sim10$ times fainter than for the bright
interior regions. The bottom panel of Figure~\ref{fig_chandra} shows
that this latter component extends right to the edge of the radio
plateau around most parts of the SNR, and is similarly sharp- and
straight-edged in most places.  In the south-eastern quadrant of the
SNR, the X-rays at the radio perimeter are much fainter than in the
rest of the source, but still extend out to this sharp outer edge.
(Note that at its most southerly and western extents, the SNR extends
beyond the extent of the {\em Chandra}\ ACIS-S CCD, so that we cannot
make statements about the outermost extent of the X-rays in these
regions.) The arrows in the upper two panels of
Figure~\ref{fig_chandra} also indicate that the radio filamentary
structures seen to the north and south of the core all have  X-ray
counterparts in the emission as seen by {\em Chandra}, a point
discussed further in \S\ref{sec_fil} below.  The analyses of Park
\etal\ (2002\nocite{prh+02}) and of Gonzalez \& Safi-Harb
(2003\nocite{gs03}) demonstrate that these filaments have thermal X-ray
spectra.

\begin{figure}[hbt]
\centerline{\psfig{file=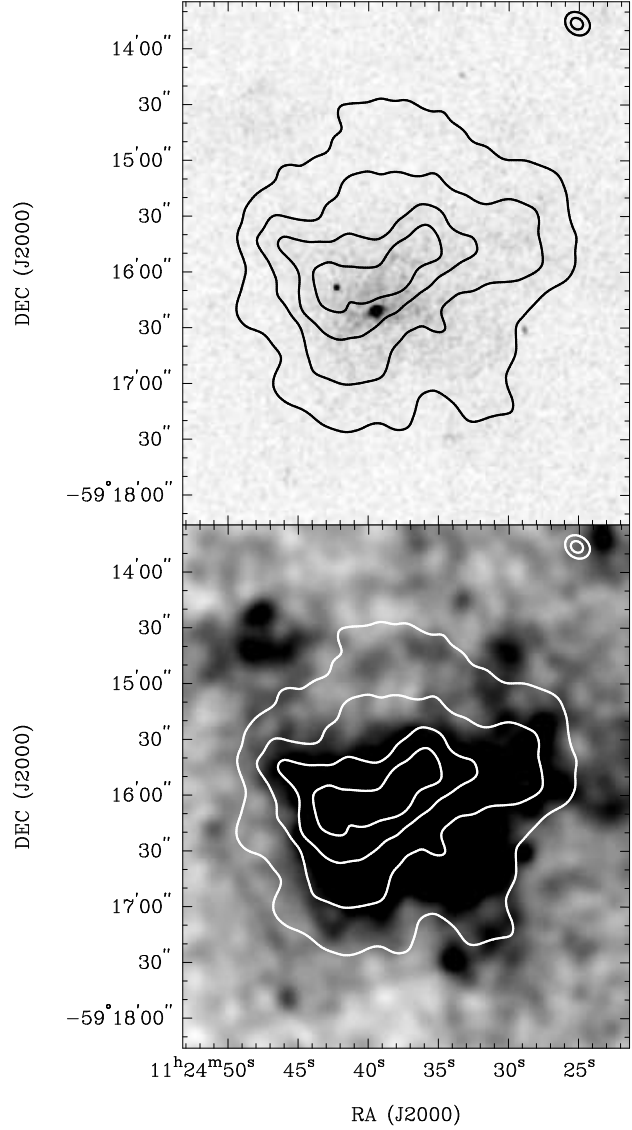,height=15cm,angle=270}}
\caption{Comparison of X-ray and radio morphologies of the central
region of SNR~\snr. In both panels, the greyscale shows the
\cxo\ ACIS-S data of Park
\etal\ (2002\protect\nocite{prh+02}) in the energy range
2.5--10~keV, while the contours show 20-cm ATCA data
at the levels of 10, 20, 30 and 40~mJy~beam$^{-1}$. 
In the upper panel, the X-ray data have been smoothed with
a $2''$ gaussian, while in the lower panel the data
have been smoothed with a $10''$ and are displayed
over a restricted greyscale range to bring out faint structure.}
\label{fig_chandra_hard}
\end{figure}

In Figure~\ref{fig_chandra_hard} we show the central region of hard
non-thermal X-rays, corresponding to the PWN identified
by Hughes \etal\ (2001\nocite{hsb+01}). The  upper panel of
Figure~\ref{fig_chandra_hard} demonstrates that the brightest regions of
the nebula are centrally concentrated, and have a much smaller extent
than the core region of the radio source.  However, as can be seen in
the lower panel, the faintest regions of the X-ray nebula extend to fill
a significant fraction of the radio core.

We have formed images of the SNR (not shown here) with the shortest
baselines excluded, so as to maximize sensitivity to an embedded point
source. Above our 5-$\sigma$ sensitivity level of 2.5~mJy at 20~cm, we
find no evidence for a central point source. This is consistent with
the 20-cm flux density for PSR~\psr\ of 80~$\mu$Jy determined by Camilo
\etal\ (2002\nocite{cmg+02}).

\subsection{Spectral Index Determination}
\label{sec_spec}

Spectral indices were derived from the images matched in $u-v$ coverage
as described in \S\ref{sec_obs}.
The calculation of spectral indices was carried
out through spatial tomography (for discussion and implementation
of this approach see \cite{kr97}; \cite{cgk+01};
\cite{dkrd02}).  Spatial tomography
between two images, $I_1$ and $I_2$, involves scaling $I_2$ by a trial
spectral index $\alpha_t$ (where $I_\nu \propto \nu^\alpha$), and then
subtracting this scaled image from $I_1$.  This then forms a difference
image,
\begin{equation} 
I_t = I_1 - \left( \frac{\lambda_2}{\lambda_1} \right)^{\alpha_t} I_2,
\label{eqn_spec}
\end{equation} 
where $\lambda_1$ and $\lambda_2$ are the observing wavelengths
corresponding to $I_1$ and $I_2$ respectively,
The spectral index, $\alpha$, of a feature is then the value of
$\alpha_t$ at which the emission from this feature blends into the
background. An uncertainty in $\alpha$ is determined by finding the
range in values of $\alpha_t$ at which the residual at this position in
the difference image becomes significant.

\begin{figure}[hbt]
\centerline{\psfig{file=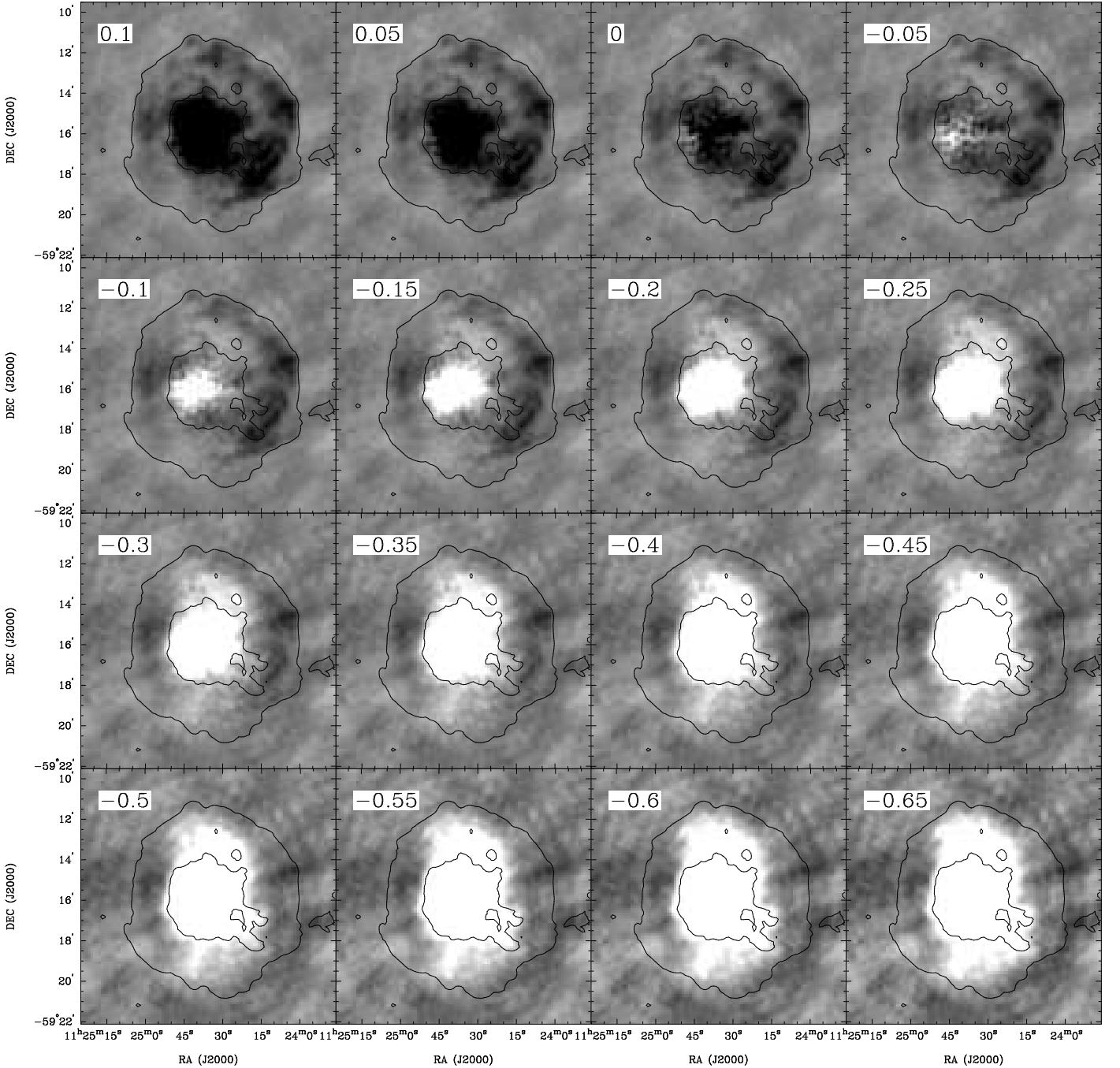,height=18cm,angle=270}}
\caption{Spectral tomography images for SNR~\snr. The greyscale
shows a series of difference images between 20 and 6~cm data
which have been matched in $u-v$ coverage. For each panel,
the trial spectral index, $\alpha_t$, is shown at upper left. The data
have been smoothed to a resolution of $15'' \times 15''$,
and the greyscale
range in each panel is --5 to +5~mJy~beam$^{-1}$. The
compact source at position  (J2000)
RA $11^{\rm h}24^{\rm m}25\fs19$, Dec $-59^{\circ}13'46\farcs4$
was subtracted 
from the 20 and 6~cm images before comparison. Overlaid are
contours from the 20~cm image in Figure~\ref{fig_all},
at the levels of 0.5 and 5~mJy~beam$^{-1}$.}
\label{fig_spec}
\end{figure}

The resulting tomography series for 20~cm vs 6~cm data is shown in
Figure~\ref{fig_spec}; comparisons between 20/13~cm and 13/6~cm are not
shown here, but give similar results. These
images clearly demonstrate spatial variations in the spectral index
distribution. We first consider the two main components of emission:
the central core and the surrounding plateau. The
 core clearly has a spectral index $\alpha = -0.05 \pm 0.05$; the
mottled appearance of the core in these panels is most likely due to
slight differences in the deconvolution process between 20 and 6~cm.
The plateau appears to have a spectral index $\alpha = -0.5 \pm
0.1$.  This can most clearly be seen along the western rim of the SNR,
where positive (black) emission can be seen in the panel corresponding
to $\alpha_t = -0.40$, and negative (white) emission can be seen at
$\alpha_t=-0.60$.

The smaller-scale structures superimposed on the plateau all have
flatter spectra than their surroundings. This can most clearly be seen
in the panels $\alpha_t =-0.3$, where emission in many regions is
white, but the outer edges of the plateau are still black.
Specifically, the filaments running north from the center of the core
have $\alpha \approx -0.05$, while that to the south has $\alpha
\approx 0$. The prominent spur extending to the south-west of the core
has a spectrum $\alpha \approx -0.1$.

\subsection{Polarimetry}

\begin{figure}[hbt]
\centerline{\psfig{file=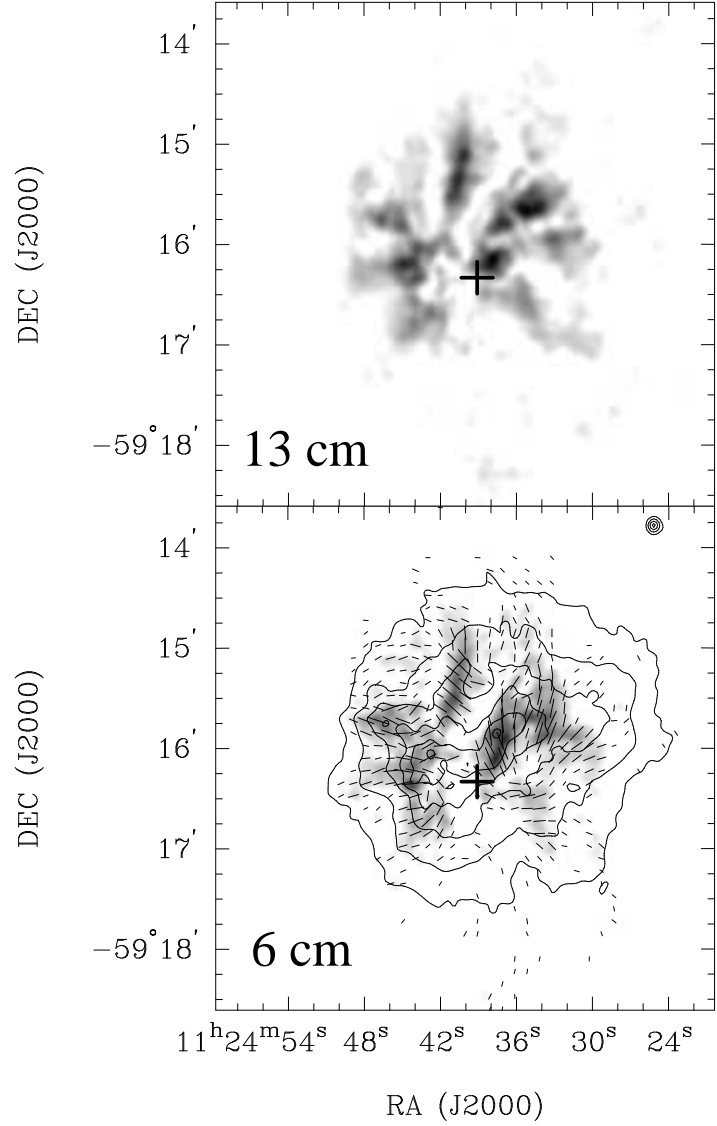,height=15cm}}
\caption{Linearly polarized emission from SNR~\snr. The upper and
lower panels 
show polarized intensity at 13 and 6~cm respectively. In the
lower panel, the image is
overlaid with vectors whose length is
proportional to the 6-cm polarized intensity, and whose orientation
indicates the magnetic field direction after correction for Faraday
rotation. The contours are 6-cm total intensity, at the levels of 2, 4,
7, 10, 13 and 16~mJy~beam$^{-1}$. In both panels the position of
PSR~\psr\ is marked with a ``+'' symbol.}
\label{fig_pol}
\end{figure}

Linearly polarized emission from SNR~\snr\ is detected in all three
wavebands.  Significant levels of polarized emission are only seen from
the core region, where typical levels of fractional polarization are
1--5\% at 20~cm, 5--15\% at 13~cm and 10--20\% at 6~cm.  The
distribution of polarized emission in the core at 13 and 6~cm is shown
in Figure~\ref{fig_pol}. The morphologies at these two wavelengths are similar,
consisting of several bright ``fingers'' of polarization, with channels
of reduced polarization between them.  There are no unusual properties
of the polarized emission at or near the position of PSR~\psr. The
20~cm polarization (not shown here) shows a comparable morphology, but
appears to suffer from significant beam and bandwidth depolarization.

In the lower panel of Figure~\ref{fig_pol} we have plotted the orientation
of the magnetic field within the core, assuming this to be oriented
perpendicular to the intrinsic position angles of polarization.  While the
field is well-ordered in small regions, no correlation between orientation
and polarized intensity is apparent --- in some regions the field runs
along the ``fingers'', and in other places is perpendicular to them.

It is difficult from these data to put strong constraints on the
presence of any polarized emission from the surrounding plateau
region.  At 20~cm the depolarization effects are severe, at 13~cm the
off-axis polarimetric response is very poor, and at 6~cm this region is
too faint and extended to be well-imaged. At 13~cm, we roughly estimate
an upper limit on the fractional polarization of this
region of $\sim20$\%.

\subsection{\HI\ Absorption}

In the top panel of Figure~\ref{fig_hi}, we show an \HI\ emission
spectrum from a region adjacent to the SNR, taken from the Southern
Galactic Plane Survey (\cite{mdgg02}).  This profile shows at least
four peaks in \HI\ emission in this direction, at LSR velocities of
approximately --30, --20, --8 and 0~\kms. In the lower panel of
Figure~\ref{fig_hi}, we show an \HI\ absorption spectrum for the bright
core of SNR~\snr.  The SNR shows clear absorption towards all four of
the peaks seen in \HI\ emission.

\section{Discussion}
\label{sec_discuss}

The nature of the radio morphology of \snr\ has been uncertain:
Lockhart \etal\ (1977\nocite{lgcm77}) argued that the whole source was
a synchrotron  nebula powered by a pulsar, while Braun
\etal\ (1986\nocite{bgcr86}) developed a model in which \snr\ had no
pulsar-powered component, but rather was a ``shell''-type SNR
interacting with a molecular cloud.  We here discuss our new data on
this source, and interpret them in the context of the recent
\cxo\ observations and pulsar detection.

\subsection{The Distance to \snr}

The \HI\ absorption profile obtained for SNR~\snr\ by Caswell
\etal\ (1975\nocite{cmr+75}) showed strong absorption at --30~\kms,
with three other possible absorption features seen in the range --30 to
0~\kms\ at lower significance.  Caswell \etal (1975\nocite{cmr+75})
interpreted this spectrum as indicating that the SNR was beyond the
tangent point in this direction. Using the Galactic rotation curves
available at that epoch, they adopted a lower limit on the distance to
\snr\ of 3.7~kpc.\ However, we note that the most
commonly quoted distance in the literature
is $4.8\pm1.6$~kpc, which was derived by Saken, Fesen \& Shull
(1992\nocite{sfs92}) using the data of Caswell \etal\ (1975\nocite{cmr+75}),
but adopting 0~\kms\ as an upper limit on the SNR's
systemic velocity.  However, such an upper limit can only be
established by a clear {\em lack}\ of \HI\ absorption towards an
emission feature (e.g.\ \cite{fw90}).  Since \HI\ emission at
0~\kms\ showed a possible counterpart in absorption, no upper limit on
the SNR's distance can be derived from these data.

\begin{figure}[hbt]
\centerline{\psfig{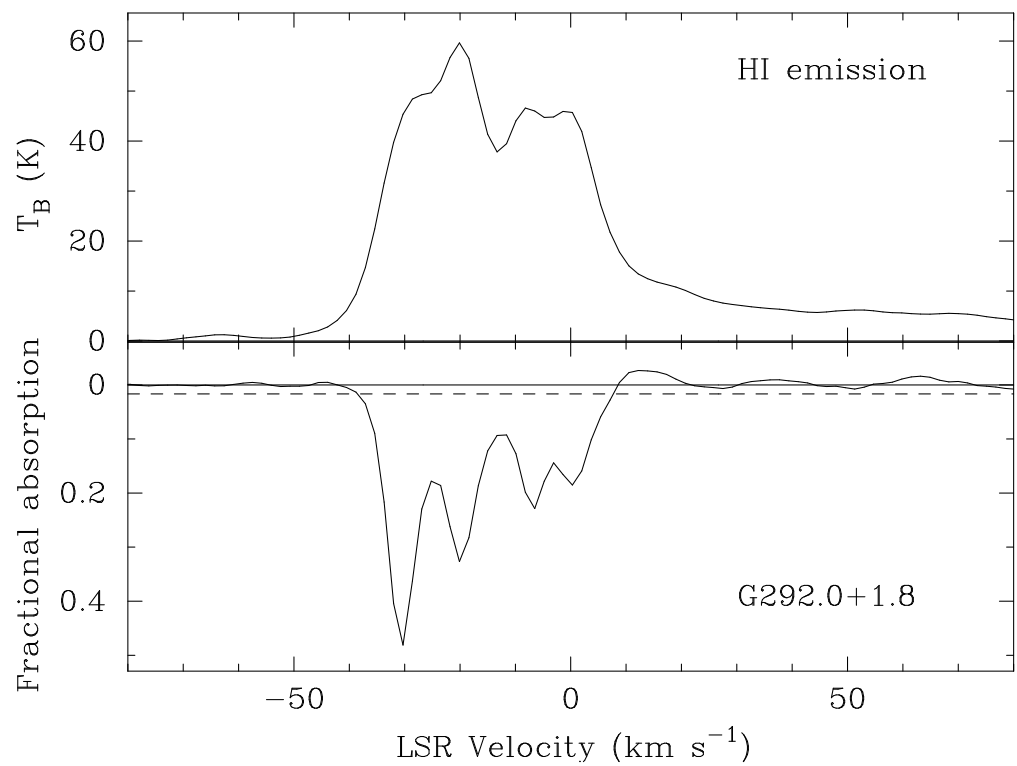}}
\caption{\HI\ emission and absorption towards SNR~\snr. The upper panel
shows the \HI\ emission profile in this region, as measured by the
Parkes component of the Southern Galactic Plane Survey
(\protect\cite{mgd+01}).  The lower panel shows \HI\ absorption towards
the central core of \snr.  In the lower panel, a dashed line marks
absorption at the 6-$\sigma$ level, where $\sigma$ is calculated from
the emission in line-free channels.}
\label{fig_hi}
\end{figure}

The absorption towards the SNR shown in Figure~\ref{fig_hi} confirms that
all four absorption lines seen by Caswell \etal\ (1975\nocite{cmr+75})
are genuine.  A formal lower limit on the SNR distance comes from the fact
that \HI\ absorption is seen out to the most negative velocities for which
gas is present, representing the tangent point in this direction. Using
the best-fit model for Galactic rotation of Fich, Blitz \& Stark
(1989\nocite{fbs89}), and adopting Galactic parameters of $\Theta_0
= 220$~\kms\ and $R_0 = 8.5$~kpc (\cite{klb86}), we adopt the tanget
point distance of 3.2~kpc as a lower limit on the SNR's distance. (This
reconfirms the result of  Caswell \etal\ 1975\nocite{cmr+75}, who came
to the same conclusion but using $R_0 = 10$~kpc.)

A further constraint on the distance can be obtained
by a comparison between the \HI\ emission spectrum
seen in Figure~\ref{fig_hi} and the foreground absorption
seen in X-rays.
For an optically thin
\HI\ emission spectrum for which the brightness temperature
is $T_B$ at a velocity $v$, the corresponding hydrogen column
density corresponding to the emitting \HI\ in the velocity
range $v_1 < v < v_2$ is:
\begin{equation}
N_H = C_H \int_{v_1}^{v_2} T_b~dv,
\label{eqn_nh}
\end{equation}
where $C_H = 1.823\times10^{18}$~cm$^{-2}$. Integrating the
\HI\ emission spectrum seen in the top panel of Figure~\ref{fig_hi}
over the velocity range $-100 < v < 0$~\kms, we correspondingly infer a
total column density  $N_H \approx 3.3 \times 10^{21}$~cm$^{-2}$ for
all \HI\ gas along the line-of-sight with negative
velocities.\footnote{Repeating the integration over the velocity range
$0 < v < 200$~\kms, we find a contribution $N_H \sim 1.8 \times
10^{21}$~cm$^{-2}$ for gas at positive velocities.} This value agrees
well with the column density $N_H =
(3-5)\times10^{21}$~cm$^{-2}$ determined from photoelectric
absorption seen in the X-ray spectrum of the  SNR~\snr\
(\cite{hsb+01}; \cite{gs03}).  
We thus conclude that all gas at negative velocities
is in front of the SNR. Independent evidence for this
conclusion is provided by the good match between
the SNR's \HI\ absorption profile and the corresponding \HI\ emission
spectrum, which together show four peaks/troughs at matching velocities.

Considering the Galactic spiral structure in this direction, we find
that at distances between $\sim2$ and $\sim8$~kpc, this line-of-sight
runs largely tangent to the Sagittarius-Carina spiral arm, of which the
Galactic rotation curve implies that gas out to 6.2~kpc will have
negative velocities, and material beyond this point will have positive
velocities. We have determined above that the SNR is behind all the
negative-velocity gas in this direction. Since it is highly unlikely
that this foreground gas is confined to only the near part of the
spiral arm, we conclude that the most likely lower limit
for the SNR is a systemic
velocity of 0~\kms, corresponding to a distance of at least
$6.2\pm0.9$~kpc.\footnote{The error quoted in the distance incorporates
both a 10\% uncertainty in comparing $N_H$ from Equation~(\ref{eqn_nh})
with that from X-rays, and a typical uncertainty of $\pm7$~\kms\ in
converting systemic velocities to distances due to the random motions
of \HI\ clouds.} This is broadly consistent with the distance to the SNR of
5.4~kpc implied by reddening of its optical filaments (\cite{gsz+79}), and with
the distance of $6.4\pm1.3$~kpc implied by the dispersion measure of
the associated pulsar \psr\ (\cite{cl02}). In future discussion, we
adopt a distance to the system of $6d_6$~kpc.

\subsection{The Core}
\label{sec_core}

The core is centrally brightened, significantly polarized,
has a flat spectrum, and contains the young pulsar~\psr.
By analogy with the many other sources known to have such properties
(see \cite{gae01b}),
we interpret the radio core as a PWN powered by PSR~\psr.

The effects of Faraday rotation, both interior to the source and in the
foreground, can produce complicated structures in polarization
(\cite{gdm+00}).  However, the morphology and intensity of such
structures show a very strong wavelength dependence. In contrast, the
structure of the polarized intensity in the core of \snr\ shows little
difference between 6 and 13~cm, which implies that Faraday effects are
negligible, and that the observed distribution of linear polarization
is intrinsic to the source. In this case, we interpret the fingers of
bright polarization as representing regions of well-ordered magnetic
field, and the linear channels of reduced polarization as resulting
from beam depolarization, where the intrinsic orientation of magnetic
field changes on scales much smaller than the resolution limit.  These
results demonstrate a complicated and tangled geometry for the nebular
magnetic field in some regions, in contrast to the well-ordered field
structure seen over the entire extent of the Crab Nebula (\cite{hv90};
\cite{hss+95}).  In \S\ref{sec_plateau} below, we argue that the PWN
is in the early stages of an interaction
with the SNR reverse shock; it is possible that this process
might produce this
complicated magnetic field geometry.

At 20~cm the PWN has a flux density of 5.5~Jy and a spectral
index $\alpha \approx -0.05$, while at 1~keV, its flux density is
$\sim2-4$~$\mu$Jy and the spectral index is $-1.1 \la \alpha \la -0.7$
(\cite{hsb+01}). Extrapolating between these two data-points, we
find that a break in the spectrum $\Delta\alpha \sim 1$ is inferred
at a frequency $\nu_b \la 800$~GHz. If we interpret such a break as
being due to synchrotron cooling, then the age
of the source allows us to compute the nebular magnetic
field (e.g. \cite{gak+02}). 

The exact age of SNR~\snr\ is unclear: model dependent analyses of the
optical and X-ray spectra emission from this source suggest ages of $t =
2000d_6$~yr (\cite{mc79}; \cite{bgdb83}) and $t = 2600$~yr (\cite{gs03})
respectively, while the pulsar's characteristic age is $\tau_c \equiv
P/2\dot{P} = 2900$~yr (\cite{cmg+02}).  If in subsequent discussion we
adopt an age $t = 2500t_1$~yr, the implied nebular magnetic field is $B_n
\ga 410 t_1^{-2/3}$~$\mu$G (e.g.\ \cite{fggd96}).  Such a large magnetic
field is difficult to explain, as the resulting synchrotron lifetime in
X-rays would then be only $\sim5$~yrs, which is inconsistent with the
observed large extent of the X-ray PWN.  A nebular magnetic field $B_n
\sim 3$~$\mu$G, as inferred from the radius of the termination shock
surrounding the pulsar, seems more likely (\cite{hs03}).  We therefore
think it likely that the implied break in the nebular spectrum is not
simply due to synchrotron losses.

A second possibility to account for the broad-band spectrum is that the
radio and X-ray emission in this source originate from entirely
distinct populations, and that to extrapolate between the flux
densities of the two regimes is not meaningful. Indeed in the Crab
Nebula, models cannot simultaneously account for both the radio and
high-energy emission from the same particle population (\cite{kc84}).

Alternatively, for a single emitting population, a PWN spectrum with
just one synchrotron break only occurs if the pulsar's rate of energy
output has been approximately constant over its lifetime (\cite{ps73};
\cite{rc84}). This is a valid assumption only at times $t < \tau_0 \equiv
2\tau_c/(n-1) - t_p$, where $t_p$ is the pulsar's true age, and $\tau_c
\equiv P/2\dot{P}$ is its characteristic age 
(\cite{ps73}).  However, for times $t > \tau_0$ (``phase 3'' of
\cite{ps73}), the spin-down luminosity of the pulsar is substantially
reduced, and the emission from the PWN effectively has contributions
from two different populations of particles. There can thus be multiple
spectral breaks in the PWN spectrum, and the simple arguments made
above then do not apply. In the case of PSR~\psr, we have that $\tau_c
= 2900$~yrs and $t_p = 2500t_1$~yr. Requiring that
$t_p > \tau_0$, we find that $n > \tau_c/t_p + 1 \ga 2.2$.  While the
braking index has not yet been measured for this pulsar, it is
certainly reasonable that it falls in this range (e.g.\ \cite{mdn85};
\cite{ckl+00}), 
in which case
multiple breaks would indeed be expected in the nebular spectrum.

Finally, it is important to note that a variety of PWNe have been
shown to have a single steep spectral break, $\Delta\alpha > 0.5$, at
a relatively low frequency, $\nu_b \la 100$~GHz.  The nature of these
low-frequency breaks is not understood (\cite{gs92}; \cite{wspb97}).
It has been speculated that they result from the anomalous properties of
the central pulsar, but the recent detection of the apparently Crab-like
young pulsar J0205+6449 in the prototypical low-frequency-break PWN
3C~58 (\cite{mss+02}) suggests that some other effect is at work.
The PWN powered by PSR~\psr\ could similarly show such properties,
and this could also explain its X-ray and radio flux densities.

The PWN in SNR~\snr\ is notable in that, as shown in
Figure~\ref{fig_chandra_hard}, its overall radio and X-ray extents are
similar. This is contrary to the situation seen in the Crab Nebula, where
the X-ray nebula is much smaller than that seen in the radio, reflecting
the much shorter synchrotron lifetimes of X-ray emitting particles.
However, in this case the very low nebular magnetic field, $B_n \sim
3$~$\mu$G (\cite{hs03}), implies a synchrotron lifetime in the X-ray band
larger than the age of the system.  X-ray emitting electrons can thus flow
out to largely fill the volume delineated by the radio PWN, resulting in
little change of the nebular extent with observing wavelength. A similar
situation is seen for the PWN powered by PSR~B1509--58, which also has a
low nebular magnetic field and hence a large X-ray extent (\cite{gak+02}).

\subsection{The Plateau}
\label{sec_plateau}

The plateau region is morphologically and spectrally distinct from the
core. However, it could be argued that this outer component is also
powered by the pulsar. In this case, the steeper spectrum of this
region could be due to synchrotron losses in particles which have
traveled further from the pulsar, as is seen in X-rays for several
other PWNe (\cite{scs+00}; \cite{bwm+01}). However, in such a case one
would expect the PWN to steadily fade and steepen in its spectrum with
increasing radius due to synchrotron losses, at odds with the sharp
transition in surface brightness and spectral index seen here.
Furthermore, for synchrotron cooling to be efficient at radio
frequencies, the nebular magnetic field would have to be even higher
than the problematic value considered in \S\ref{sec_core} above.

Given that the X-ray PWN coincides with the radio PWN, but that thermal
X-ray emission extends over most of the radio plateau region, we
therefore interpret the plateau as corresponding to the standard
``shell'' component of the SNR, representing the supernova shock and
its interaction with its environment. Given this interpretation,
\snr\ clearly can be classified as a ``composite'' SNR, showing a
characteristic flat-spectrum core surrounded by a steep-spectrum shell
(e.g.\ \cite{hb87}).  In this sense, \snr, along with G11.2--0.3,
Kes~75 and 0540--69.3, is one of a handful of ``textbook'' SNRs which
show a central young pulsar, its associated X-ray/radio synchrotron
nebula, and an enveloping SNR shell.

Figure~\ref{fig_profile} demonstrates that the plateau shows a sharp
outer edge, while the lower panel of Figure~\ref{fig_chandra} shows
that this edge also corresponds in most places to a faint and similarly
sharp-edged perimeter of X-ray emission. Several other young SNRs are
sharp-rimmed in both radio and X-rays, most notably SN~1006, Tycho's
SNR and Cas~A (\cite{rg86}; \cite{dvs91}; \cite{gkr+01}; \cite{hdhp02}).
This morphology is thought to indicate that diffusive shock acceleration
is occurring at the outer edge of the SNR --- the sharpness of the edge
is due to enhanced turbulence ahead of the SNR blast wave, which prevents
synchrotron-emitting electrons from diffusing upstream (\cite{abr94}).
Thus, as for other sharp-rimmed SNRs, we identify this outer radio
and X-ray edge as corresponding to the forward shock of the SNR, where
ambient gas is first shocked by the remnant.  The radius of this shock
is $4'$, corresponding to a shock radius $R = 7.0d_6$~pc. 
At least in the case of SN~1006,
the sharp outer edge of X-ray emission has been
shown to be non-thermal, corresponding to electrons which have been
accelerated by the shock up to TeV energies. In the case of SNR~\snr,
the available data suggest a thermal spectrum for this outermost rim
(\cite{prh+02}; \cite{gs03}).  However, this region is very faint, and
deeper X-ray observations will be needed to constrain the presence of
any non-thermal component.
Park \etal\ (2002\nocite{prh+02}) identify a continuous
ring of X-ray filaments $1'-2'$ interior to this outer edge, and
propose that it corresponds either to the SNR forward shock or to
circumstellar material overrun by the shock. Clearly the fact that
both radio and X-ray emission from the SNR extend beyond this
filamentary ring favors the latter alternative.

The identification of a complete outer rim representing the
forward shock allows us to estimate the position of the SNR's
geometric center. We fit a circle to this outer rim,
and find a best-fit center
of (J2000) RA $11^{\rm h}24^{\rm m}34\fs8$, Dec $-59^{\circ}15'52\farcs9$,
with an uncertainty of $\pm5''$ in each coordinate. The position
of PSR~\psr\ as determined by Hughes \etal\ (2001\nocite{hsb+01}) is
offset from this position by $\approx45''$. If the
SNR's geometric center coincides with the explosion
site, we can infer a projected
pulsar velocity for PSR~\psr\ of $520d_6/t_1$~\kms, which is
typical of the velocity distribution seen for the Galactic
pulsar population (\cite{acc02}). Despite this high pulsar velocity,
the PWN does not have a bow-shock morphology because its motion
is currently not supersonic in the gas in the SNR's interior
(\cite{vag98}).

An important difference between \snr\ and the vast majority of other
SNRs is that most SNRs show significant limb-brightening, with little
interior radio emission other than that corresponding to an associated PWN.
This appearance is thought to correspond to a thin spherical shell of
synchrotron emission, which we only see around the edges because of
projection effects. However, in the case of \snr, little or no
limb-brightening is seen around the perimeter of the plateau, except
perhaps along the south-western edge. Rather, the plateau region
of \snr\ appears as a very thick annulus, with a ratio of outer
radius to inner radius of $\sim2$. 

Once an expanding SNR has swept up a significant amount of mass from the
ambient medium, it enters the ``Sedov-Taylor'' phase of evolution during
which it takes on a structure consisting of both the forward blast wave
and a reverse shock interacting with supernova ejecta.  We expect radio
emission to be produced not just at the outer blast wave, but also through
turbulence and particle acceleration in relativistic material between the
two shocks (e.g.\ \cite{jj99}).  Thus the interpretation we propose for
the thickness of the plateau region is that SNR~\snr\ is in a moderately
evolved state, in which the forward and
reverse shocks are at markedly different radii,
and in which radio emission fills the
region between them. Well-separated forward
and reverse shocks with a thick
radio plateau between them have been clearly identified in Cas~A
(\cite{arl+91}; \cite{gkr+01}), while thick radio
shells as seen here have been similarly interpreted in SNRs Kes~79
(G33.6+0.1) and DA~495 (G65.7+1.2) (\cite{vbgh89}; \cite{vbs91}).
Indeed, from earlier radio data on \snr\ in which the plateau and core
regions were not identified as being spectrally distinct, Braun \etal\
(1986\nocite{bgcr86}) interpreted the entirety of \snr\ as corresponding
to part of such a thick shell.

Truelove \& McKee (1999\nocite{tm99a}, hereafter TM99) present detailed
models for the evolution of a SNR and its shock structure. The exact
positions of the forward and reverse shocks depend on the density
distributions of the ejecta and of the ambient medium, but for a wide
variety of such possibilities, TM99\nocite{tm99a} show that a ratio
of forward shock to reverse shock radius of $\sim2$ typically occurs
after a time $t \sim 1.4 t_{ch}$, at which point the radius of the
forward shock is $R \sim 1.25 R_{ch}$ and its velocity is $V \sim 0.4
V_{ch}$, where $t_{ch}$ and $R_{ch}$ are characteristic time and size
scales for the system, respectively, and $V_{ch} = R_{ch}/t_{ch}$.
Since $R = 7.0d_6$~pc and $t = 2500t_1$~yr, we can thus infer that
$R_{ch} \approx 5.6d_6$~pc, $t_{ch} \approx 1800t_1$~yr and $V_{ch}
= 3100d_6/t_1$~\kms. The latter implies that $V \sim 0.4V_{ch} \approx
1200d_6/t_1$~\kms\ which is broadly consistent with the
expansion velocity inferred from
optical spectroscopy (\cite{bgdb83}).  These filaments are coincident
with the PWN in projection, but are presumably physically located near
the front or back of the SNR. The rate of expansion implied by the
inferred velocity is $43/t_1$~mas~yr$^{-1}$, which should be
detectable with both {\em Chandra}\ and the ATCA in another $\sim5$~years.

Assuming that the medium into which the SNR expands is homogeneous,
Table~2 of TM99 gives that:
\begin{equation}
R_{ch} = 3.07 M_{ej}^{1/3} n_0^{-1/3}~{\rm pc}
\label{eqn_rch}
\end{equation}
and
\begin{equation}
t_{ch} = 423 E_{51}^{-1/2} M_{ej}^{5/6} n_0^{-1/3}~{\rm yrs},
\label{eqn_tch}
\end{equation}
where $M_{ej}~M_\odot$ is the mass of the supernova ejecta,
$n_0$~cm$^{-3}$ is the atomic density of the ambient medium and $E_{SN} = E_{51}
\times 10^{51}$~erg is the kinetic energy of the supernova explosion.  
With these
definitions, it can be shown that the ratio of swept-up mass,
$M_{sw}$, to ejected
mass is $X \equiv M_{sw}/M_{ej} \approx 3.0 (R/R_{ch})^3$. Thus in the
case of \snr, independently of distance we can infer $X \sim 6$.  
Solving Equations~(\ref{eqn_rch}) and~(\ref{eqn_tch}) simultaneously
to eliminate $M_{ej}$ and $n_0$ respectively, 
we find that:
\begin{equation}
n_0 = 0.9 d_6^{-5} t_1^2 E_{51}~{\rm cm}^{-3}
\label{eqn_n0}
\end{equation}
and
\begin{equation}
M_{ej} = 5.4 d_6^{-2} t_1^2 E_{51}~M_\odot.
\label{eqn_mej}
\end{equation}

Braun \etal\ (1986\nocite{bgcr86}) interpret the infrared emission
from the SNR as resulting from shock-heated dust. Re-arranging 
their Equation~(4), one obtains:
\begin{equation}
n_0 = 0.7 \left(\frac{V}{1000~{\rm km~s}^{-1}}\right)^{-5/3} \left(\frac{\lambda_{peak}}
{40~\mu m}\right)^{-50/9}~{\rm cm}^{-3},
\label{eqn_ir}
\end{equation}
where $\lambda_{peak}$ is the peak wavelength of infrared emission.
Braun \etal\ (1986\nocite{bgcr86}) determine $\lambda_{peak} =
36\pm1$~$\mu$m; Saken, Fesen \& Shull (1992\nocite{sfs92}) obtain the
same value in their analysis. For $V=1200d_6/t_1$~\kms\ as determined
above, Equation~(\ref{eqn_ir})
then yields $n_0 \approx 0.9d_6^{-5/3}t_1^{5/3}$~cm$^{-3}$.
Substituting this density into of this value into 
Equations (\ref{eqn_n0})
and (\ref{eqn_mej}), we find that
$E_{SN} \approx
1.1 d_6^{10/3} t_1^{-1/3}\times10^{51}$~erg and $M_{ej} \approx 5.9 d_6^{4/3}
t_1^{5/3}$~$M_\odot$, both typical values for a core-collapse SNR.

An important point to note is that the above calculations assume that
the reverse shock progresses steadily through supernova ejecta.
This is a reasonable assumption in the case of a shell SNR,
but the presence of a central PWN complicates the situation ---
the expanding PWN will eventually collide with the reverse shock,
causing a complicated intermediate phase in the system's evolution
(\cite{rc84}; \cite{che98}). Specifically, the boundary between the SNR
reverse shock and the PWN's forward shock will reverberate back and forth
until pressure balance is established, crushing the PWN in the process
(\cite{vagt01}; \cite{bcf01}).  Clearly if the reverse shock's position
has been substantially affected by this process, its position can then
not be compared to that in standard SNR evolution models.  The fact that
there is no obvious gap between the inner boundary of the plateau and
the outer regions of the core suggests that the reverse shock and the
PWN may have indeed already collided and interacted.

However, in this case we think it likely that the reverse shock has
only recently begun interacting
with the PWN, and that the SNR evolution has not
yet been significantly affected by this process.  Our main argument for
this is simply that the PWN radius is larger than would be expected
during this interaction process.  Equation~(5) of Blondin
\etal\ (2001\nocite{bcf01}) provides the radius of the PWN, $R_p$, as a
function of time before it collides with the SNR reverse shock. For the
parameters corresponding to \snr\ and PSR~\psr, the
model of Blondin
\etal\ (2001\nocite{bcf01}) predicts $R_p \sim
3$~pc, in approximate agreement with the value $R_p \approx 3.5d_6$~pc
observed for the radio core.
However, during the subsequent interaction process, the PWN
is typically 2--5 times smaller than this (\cite{vagt01}; \cite{bcf01}).
At later times, when
pressure balance has been established, the PWN then expands
steadily again. However, in this stage the compression of the nebula
results in a small X-ray PWN centered on the pulsar, substantially offset
from a larger radio PWN with an irregular morphology
(\cite{che98}; \cite{bcf01}).
This is not observed in the case of
\snr, for which Figures~\ref{fig_all} and \ref{fig_chandra_hard} 
demonstrate the radio morphology to
be broadly circular, the pulsar to be centrally located,
and the radio and X-ray emission from the PWN to roughly correspond.
In summary, we conclude that in this case the reverse shock is only
now beginning to interact with the PWN, and that the SNR evolution
model of TM99 can still be validly applied.

\subsection{Flat Spectrum Filaments}
\label{sec_fil}

We demonstrated in \S\ref{sec_spec} that many of the filamentary
structures in the plateau region have flat spectra, $-0.1 \le \alpha \le 0$,
in contrast to the steeper spectrum for the diffuse plateau emission,
$\alpha = -0.50$. Since these filaments all have counterparts
seen in thermal X-ray emission with {\em Chandra}, 
we can thus rule out that these flat-spectrum structures are part of
the PWN.  It rather appears that there are two discrete contributions
to the shell component of the SNR, the plateau and the filaments,
distinguished by their distinct spectral indices.

While significant variations in spectral index have been seen for many
other shell SNRs (see \cite{ar93} for a review), what is striking
here is the clear morphological separation between the steep-spectrum
diffuse emission, and the flat-spectrum filaments. Only in a few other
SNRs, most notably Cas~A and the Cygnus Loop, is there such a clear
correlation between the structure of emitting features and their
spectral indices (\cite{ar96}; \cite{lr98}).

It is not easy to produce the flat spectral index seen for the
filaments via the same diffuse shock acceleration process as we have
claimed is operating for the steeper spectrum plateau. For an adiabatic
shock in the test-particle approximation, the maximum compression ratio
is 4, and the corresponding synchrotron spectral index of accelerated
particles will be no flatter than $\alpha = -0.5$. Non-linear processes
such as the escape of high energy particles from the shock, the
presence of relativistic particles with their softer equation-of-state,
and the smoothing of the shock via the back-pressure of accelerated
particles can all serve to result in higher effective compression
ratios and thus flatter emergent synchrotron spectra. However, when
these effects are considered in detail, the resultant spectral index is
no flatter than $\alpha = -0.25$ (\cite{er91}; \cite{be99}), which is
insufficient to explain the still flatter values seen here.

Another way in which a spectral index flatter than $\alpha = -0.5$ can
be produced via diffuse shock acceleration is if the particle
population encounters and is accelerated by multiple shocks.  While a
completely flat spectrum, $\alpha = 0$, can only be achieved in the
limit of an infinite number of shocks (\cite{mp93} and references
therein), only $\sim5$ successive shocks are needed to produce spectra
as flat as $\alpha \approx -0.1$ (\cite{gj00}). The turbulent and
unstable nature of the gas flow between the forward and reverse shocks
(e.g.\ Jun \& Norman 1996a,b\nocite{jn96b,jn96}) may indeed present the
opportunity for particles to be shocked and accelerated on multiple
occasions. However, it is difficult to see in such a situation how the
flat-spectrum regions could be maintained only in discrete filamentary
structures, morphologically distinct from steeper surrounding diffuse
emission. We thus regard this possibility also as unlikely.

de~Jager \& Mastichiadis (1997\nocite{dm97b}) have proposed that in the
case of SNR~W44, flat-spectrum emission in the shell is produced via
electron injection into the shell by the associated pulsar PSR~B1853+01.
Applying equipartition energy arguments to the synchrotron emission from
the plateau component of \snr\ (\cite{pac70}), we find for this region
$E_{min} \approx 5\times10^{48}$~erg.  The energy lost by PSR~\psr\
since birth is $\frac{1}{2} I (\omega_0^2 - \omega^2)$, where 
$I \equiv 10^{45}$~g~cm$^2$, $\omega =
2\pi/P \approx 47$~s$^{-1}$, and $\omega_0$ is the initial value of $\omega$.
For sufficiently
rapid initial periods ($P_0 = 2\pi/\omega_0 \la 20$~ms), 
it is energetically feasible
for the pulsar to have both powered its PWN (which
has an internal energy $E_{PWN} \approx 4\times10^{49}$~erg;
\cite{hs03}) and have provided significant energy to the
plateau region as well.  However, while we cannot completely rule out some level
of mixing between particles in the PWN and the shell, particularly once
the reverse-shock interaction discussed in \S\ref{sec_plateau} begins, it
seems unlikely that this process would occur only in the select regions
corresponding to the flat-spectrum filaments, particularly given the
broadly circular appearance of the PWN.

Leahy \& Roger (1998\nocite{lr98}) identify a collection of
flat-spectrum filaments in the Cygnus Loop, potentially similar to the
structures seen here for \snr. Leahy \& Roger (1998\nocite{lr98}) argue
that in their case, this flattening is caused by free-free absorption
by thermal material within the filaments. Since the flat-spectrum
structures seen in \snr\ coincide with X-ray emitting filaments, we can
use the X-ray properties of this source to estimate whether such an
effect is likely to be operating here. Gonzalez \& Safi-Harb
(2003\nocite{gs03}) have carried out spectral fits to a variety
of regions within the SNR using high-resolution {\em Chandra}\ data.
The flat-spectrum radio filaments correspond to their regions
16, 17 and 18, for which the X-ray spectra imply 
electron temperatures $T_e \approx 1 \times 10^7$~K
and volume emission measures $n_e^2 V_e \approx  3d_6^2\times10^{56}$~cm$^{-3}$
(Safi-Harb, private communication, 2003),
where $n_e$ is the electron density of
emitting material, and $V_e$ is the emitting volume.
Assuming a depth comparable to their extent on the sky,
the emitting volumes of each of these regions
is $V_e \sim 2d_6^3\times10^{55}$, so that $n_e \sim 4d_6^{-1/2}$~cm$^{-2}$.
For the inferred values of $T_e$ and $n_e$ the resulting
optical depth for free-free absorption is completely negligible, and
this effect is thus not likely to be operating here.

An intriguing possibility is that the flat- and steep-spectrum components
represent two distinct mechanisms for particle acceleration.  It seems
clear from the discussion in \S\ref{sec_plateau} above that diffuse
shock acceleration is occurring at the outer edge of the radio SNR,
to which the plateau emission with $\alpha = -0.5$ can be attributed.
However, synchrotron emission in SNRs may also result from either
simple compression of an ambient population of relativistic particles
(\cite{laa62a}), or from stochastic (second-order Fermi) particle
acceleration in shocked gas (\cite{cs84b}). We consider in turn whether
either of these mechanisms could be taking place here and could produce
the flat-spectrum filaments.

Very high compression ratios can be achieved in radiative shocks.  In
such cases, the ambient electron cosmic ray population can be
compressed to produce bright synchrotron emission, with a spectrum
distinct from that of the particle population accelerated by the shock.
Longair (1997\nocite{lon97}, p.\ 284) demonstrates that the synchrotron
spectrum produced by cosmic ray electrons can be as flat as $\alpha =
-0.3$ for electrons with energies $<100$~MeV.  However, to cause such
particles to radiate at $\sim5$~GHz requires an extreme compression of
$>10^4$, and cannot produce a sufficiently flat spectrum to match our
observations in any case.

Finally, we consider the possibility that second-order Fermi
acceleration could be operating in these filaments. The diffusion in
momentum space resulting from this process can flatten the particle
spectrum, and can result in synchrotron spectra as flat as $\alpha
\approx -0.1$ as required here (\cite{sf89}; \cite{ost99}). Such effects can
only dominate when $\beta \equiv P_{th}/P_{mag} \la 0.05$, where
$P_{th}$ and $P_{mag}$ are the thermal and magnetic pressure in the
gas, respectively (\cite{sf89}). The density and temperature in the
filaments, inferred above from the X-ray analysis of Gonzalez
\& Safi-Harb (2003\nocite{gs03}), imply a thermal pressure $P_{th} \approx
5\times10^{-9}$~erg~cm$^{-3}$.  For $\beta \la 0.05$ we thus require
$P_{mag} \ga 1\times10^{-7}$~erg~cm$^{-3}$ and hence a magnetic field
in the filaments $B\ga 1$~mG. For a typical ambient ISM magnetic field
strength of $\sim3$~$\mu$G, compression by a strong shock should
produce a magnetic field in the emitting regions of only 12~$\mu$G.
However, Jun \& Norman (1996b\nocite{jn96b}) have demonstrated that this
magnetic field can be strongly amplified around the edges of elongated
``fingers'' produced by Rayleigh-Taylor instabilities at the contact
discontinuity between the forward and reverse shocks. In their
two-dimensional simulations, these authors\nocite{jn96b} find a
magnetic field amplification by up to a factor of 60 over ambient
values. Jun \& Norman (1996b\nocite{jn96b}) 
argue that in three dimensions and at higher
resolutions, much higher field amplifications are likely to
result (see also \cite{jn96}). We thus argue that magnetic fields as
strong as required here can indeed be produced in SNRs, and that they
specifically should result in elongated radial structures lying between
the forward and reverse shocks.  We therefore interpret the
flat-spectrum filaments seen in \snr\ as Rayleigh-Taylor instabilities
at the contact discontinuity between ejected and ambient material in
the shock, where amplification has produced high
magnetic fields and hence efficient second-order Fermi acceleration.
This process produces the flat spectral index seen in the filaments.

In comparison to the other possibilities considered above, this
interpretation alone can account for the very flat spectra seen only in
these distinct, elongated filaments. The key to confirming
this interpretation lies in the identification of linearly
polarized emission from these regions, which we predict
to be a significant fraction of the total and to have
a radially aligned underlying magnetic field orientation. Sensitive
polarimetric observations of the plateau region will be needed
to investigate this possibility.

\section{Conclusions}

We have presented a detailed multi-frequency radio study of the young
oxygen-rich SNR~\snr\ and its pulsar wind nebula. We find that
the SNR morphology is composed of three main components:
a central core, a surrounding plateau, and a set of radial filaments
superimposed on the plateau.

The core is polarized, filled-center, has a flat radio spectrum ($\alpha
= -0.05\pm0.05$) and is coincident with PSR~\psr. We conclude that
this source represents radio emission produced by the PWN powered by
the pulsar. Comparison of the radio and X-ray spectra for this source
indicates that there must be a spectral break between the two bands;
this steepening cannot be accounted for by synchrotron losses.  The high
surface brightness of the core allows us to measure \HI\ absorption in
this direction at high signal-to-noise.  This spectrum shows absorption
out to the tangent point along this line-of-sight, implying a lower limit
on the SNR distance of 3.2~kpc. Comparison of the hydrogen absorbing
column implied by X-ray and \HI\ observations in this direction, along
with the shape of the \HI\ absorption profile seen towards the SNR, argue
that the SNR is behind all \HI\ within the solar circle, implying a more
likely minimum distance of 6~kpc to the system.

We interpret the steep-spectrum ($\alpha = -0.5\pm0.1$) plateau which
surrounds the core as corresponding to the SNR shell.  The plateau
shows a sharp outer edge in both radio and X-rays, indicating the
presence of diffusive shock acceleration at the forward shock. The
thickness of the plateau indicates that the SNR reverse shock is at a
much smaller radius, having significantly progressed back towards the
center of the SNR in the reference frame of the outwardly expanding
ejecta. The relative positions of the forward and reverse shocks allow
us to infer a shock velocity $\sim1200$~\kms, an ambient density
$\sim0.9$~cm$^{-3}$, an initial explosion energy
$\sim1.1\times10^{51}$~erg and an ejected mass $\sim5.9$~M$_\odot$,
assuming a distance to the source of 6~kpc and an age of 2500~yr. The
inferred location of the reverse shock suggests that this region is
beginning to interact with the PWN, but the morphology of the PWN
indicates that this process is yet to significantly affect the
evolution of the system.

Finally, we identify elongated radial filaments superimposed on the
plateau, with a much flatter spectrum ($-0.1 \la \alpha \la 0$) than
the surrounding steep-spectrum shell. We consider a number of possible
mechanisms for this flat-spectrum emission, and conclude that the most
likely explanation is that these filaments represent Rayleigh-Taylor
instabilities near the SNR contact discontinuity, where the magnetic field
is greatly enhanced and hence efficient second-order Fermi acceleration
can occur.

These new radio observations, combined with the recent {\em Chandra}\
imaging of this source and the subsequent detection of a young radio
and X-ray pulsar, result in a consistent picture in which SNR~\snr\ is
a moderately evolved SNR, resulting from the core-collapse of a massive
progenitor.  However, \snr\ is clearly an unusual and complicated
system in many respects. The thick shell with a sharp outer edge,
flat-spectrum filaments, and coincidence between the positions of the
reverse shock and outer boundary of the PWN are all phenomena seen in
few other SNRs, and certainly not all in the same source. Clearly
SNR~\snr\ is at a special point in its evolution, only seen at an age
of $\sim2000-3000$~yr, and only for systems containing both bright SNR
shells and energetic pulsars. Considering the known population of young
pulsars in SNRs (e.g.\ \cite{kh02}), it is clear that only
SNR~\snr\ and PSR~\psr\ meet these requirements.  We now look forward
to forthcoming detailed spectroscopic studies of this source with {\em
Chandra}\ and {\em XMM}, the results from which will be able to better
constrain the evolutionary state of and conditions within this
remarkable SNR.

\begin{acknowledgements}

We thank Larry Rudnick, Pat Slane, Jack Hughes and Tracey DeLaney for
a series of very helpful discussions, and Samar Safi-Harb for supplying
information on her spectral fits.  We acknowledge the efforts of Grant
Gussie in carrying out the original observations for this project.
The Australia Telescope is funded by the Commonwealth of Australia for
operation as a National Facility managed by CSIRO.

\end{acknowledgements}


\clearpage

\begin{table}[hbt]
\caption{Observations of \snr\ with the ATCA.}
\label{tab_obs}
\begin{tabular}{lcccc} \hline
\multicolumn{1}{c}{Date} & Observing   & Array   & Time on     & Secondary \\
            & Freq (MHz)  & Configuration & Source (hr) & Calibrator \\ \hline
1994 Apr 07 & 1376 / 1420$^a$ & 375           & 7.5  & \cala \\
1994 May 04 & 1376 / 1420$^a$ & 6D            & 8.2  & \cala \\
1997 Apr 14 & 2240 / 2496     & 375           & 10.7 & \calb \\
1997 Apr 17 & 5056 / 5440 & 375           & 9.9  & \calb \\
1997 Jun 21 & --- / 2240 & 6A            & 12.3 & \calb \\
1997 Sep 05 & 1344 / 1420$^a$ & 1.5C          & 12.0 & \calb \\
1997 Sep 06 & 4800 / 5312 & 1.5C          & 8.5  & \calb \\
1997 Oct 18 & 1472 / 2496 & 750C          & 9.7  & \calb \\ \hline
\end{tabular}

$^a$ Observations in the \HI\ line. \\
\end{table}

\bigskip
\begin{table}[hbt]
\caption{Observed properties of \snr.}
\label{tab_src}
\begin{tabular}{ccccccccccc} \hline
Waveband  & & \multicolumn{3}{c}{Flux Density (Jy)} & & Resolution &  & \multicolumn{2}{c}{Sensitivity 
($\mu$Jy~beam$^{-1}$)} \\
(cm)      & & Core & Plateau & Total & & (arcsec) & & Stokes~$I$ & Stokes~$V$ \\ \hline
20  & & $5.5\pm0.1$ & $6.4\pm0.1$ & $11.9\pm0.1$ & & $9.6\times8.0$ & & 120 & 30 \\
13 & & $5.6\pm0.1$ & $5.8\pm0.1$ & $11.4\pm0.1$ & & $7.2\times6.2$ & & 60 & 60 \\
6  & & $5.3\pm0.1$ & $3.5\pm0.1$ & $8.8\pm0.1$ & & $5.5\times4.8$ & & 110 & 100 \\
\hline
\end{tabular}
\end{table}

\end{document}